\documentclass{PoS}

\title{Light Detection in DUNE Dual-Phase}

\ShortTitle{Light Detection in DUNE Dual-Phase}

\author{\speaker{J. Soto-Oton}\thanks{On behalf of the DUNE Collaboration.}\\
        CIEMAT\\
        Madrid, SPAIN\\
        E-mail: \email{j.soto@cern.ch}}

\abstract{DUNE will be an underground neutrino oscillation experiment that will perform precision measurements of the PMNS mixing parameters, determine unambiguously the mass ordering and discover leptonic CP violation. It also comprises a rich non-accelerator physics program as the detection of supernova neutrinos, nucleon decay and BSM physics.
One of the modules of the DUNE is proposed to be Dual-Phase LArTPC. Inside this module, a light detection system (LDS) is being designed, consisting on an array of photomultiplier tubes and a calibration system based on optical fibers. To fulfil the physics program, the LDS is aimed to comply with certain physics requirements. Those are to provide a detection efficiency of more than 90\% for a Supernova Burst within the Milky Way and an event time reconstruction efficiency of more than 90\% with a signal purity of more than 90\% across the active volume for proton decay event candidates.
The present document summarizes the status of the simulation studies of the light detection in DUNE Dual-Phase, and the expected performance of the LDS, that will be part of the forthcoming Technical Design Report of DUNE.}

\FullConference{XXIX International Symposium on Lepton Photon Interactions at High Energies - LeptonPhoton2019\\
		August 5-10, 2019\\
		Toronto, Canada}

\begin{document}

\section{Introduction}
Proposed for construction 1.5 km underground in the Sanford Underground Research Facility (South Dakota, US), the Deep Underground Neutrino Experiment (DUNE) will consist of four 10 kton fiducial mass modules of liquid argon (LAr) to detect a beam of neutrinos sent from Fermilab, 1,300 km away. DUNE physics program comprises neutrino oscillation measurements, the mass hierarchy determination and to discover CP violation. It will be able to detect supernova neutrinos, and to perform nucleon decay searches and BSM physics [1].
\par In the first phase of DUNE, two liquid argon Time Projection Chamber (LAr TPC) modules of different technologies are proposed to be built, with equivalent prototypes being assembled now at CERN (protoDUNEs): a first single-phase module, and a second dual-phase with both liquid and gaseous argon phases to additionally amplify the signal. The dual-phase module will have a drift distance of 12 meters.
\vspace{-0.1cm}
\subsection{The importance of the light detection in DUNE}
Inside a LAr TPC ionizing particles produce primary scintillation light and ionization electrons. In a dual-phase LAr TPC, electrons are drifted in the liquid, and extracted and amplified inside the gas argon phase, before being readout, producing secondary electroluminescence light in gas. The light detection system (LDS) uses the primary scintillation light to provide the event time reconstruction. This time allows to determine the third coordinate needed to reconstruct 3 dimensional tracks, which is redundant in the case of beam events, but crucial for non-beam events. In the same way, the LDS provides a trigger for non-beam events, and it can contribute to the calorimetric reconstruction.

\vspace{-0.15cm}
\section{Light Detection System of DUNE Dual-Phase}
The baseline design of the DUNE Dual-Phase LDS consists on an array of 720 photomultiplier tubes (PMT) of 8\,'' diameter and 14 dynode chain, that are highly sensitive to visible light and placed at the bottom of the detector. Since the scintillation light is produced in the VUV range (128nm), PMTs are coated with Tetraphenyl butadiene (TPB), a highly efficient wavelength shifter with the reemission peak at 430nm [2]. The half-top part of the detector walls will be covered with TPB-coated reflective foils, to increase the amount of detected light generated in the top part of the detector.
A LED-based fiber calibration system is also proposed in order to monitor the gain of each PMT [3].

\vspace{-0.15cm}
\section{Performance of the system with simulation studies}
In order to ensure that the system fulfil the DUNE physics requirements, a detailed model of the detector has been simulated, including the generation and propagation of the scintillation light in the liquid argon, the photon detectors response and the signal reconstruction.
\par To validate the effect of the reflector foils, three geometries have been tested: No foils, foils entirely covering all detector walls (full foil), and foils covering only the upper half (half foil).
Additionally, a radiological model has been considered as a background source of light. This model includes the liquid argon radiological activity from $^{39}Ar$, $^{42}Ar$, and $^{222}Rn$ and neutrons from the underground rock.

\vspace{-0.15cm}
\subsection{Light yield and non-uniformity}
\begin{figure}
\centering
\includegraphics[width=.43\textwidth]{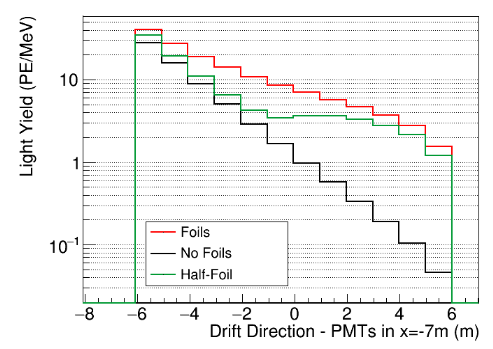}
\caption{Expected light yield in the full detector. The yield units are the number of detected photo-electrons per MeV of deposited energy. Three PD system designs are compared: No foils (black), foils (red) and half foils (green, baseline).}
\label{fig1}
\end{figure}

As the detector is conceived as a monolithic volume with all the PMTs placed at the bottom, the performance of the light detection system can be compromised in the top part of the detector,  12 meters away from the LDS. The inclusion of reflective foils attenuates the non-uniformity in the light yield, improving significantly the performance of the system detecting events produced in the top part of the detector by more than one order of magnitude. Figure \ref{fig1} shows the expected light yield as a function of the drift coordinate and averaging over the two other spatial coordinates [4].

\vspace{-0.15cm}

\vspace{-0.5cm}

\subsection{Supernovae neutrinos light trigger}
\begin{figure}
\centering
\includegraphics[width=0.75\textwidth]{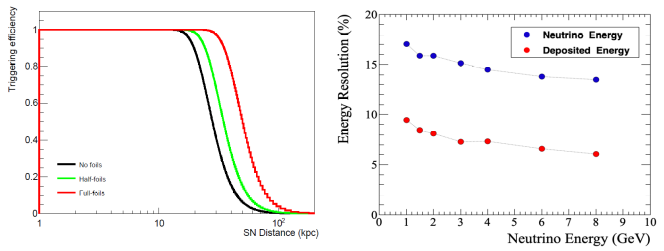}
\caption{Left Panel: Triggering efficiency of the Dual-Phase LDS to a supernova burst vs distance. Right Panel: Energy resolution of the neutrino energy (blue) and deposited energy (red) provided by the LDS as a function of the neutrino energy for $\nu_e$ CC interactions in the LAr fiducial volume.}
\label{fig2}
\end{figure}

The LDS will provide a trigger for supernova burst (SNB) within the galaxy (i.e. at 10kpc distance). A SNB in our galaxy would produce thousands of neutrinos in the range of 5-50MeV interacting with our detector. This will be detected as a jump in the rate of low energy events during a certain time window, with respect to the low energy light events produced by radiological backgrounds, mainly neutrons coming from the underground rock. Considering a time window of 2s and a fake trigger rate of 1 event per month, the light detection system can provide a 90\% triggering efficiency on a supernova burst happening at 24kpc (left panel of Figure \ref{fig2}) [4].

\vspace{-0.1cm}

\vspace{-0.2cm}

\subsection{Beam neutrino energy reconstruction}
The LDS can provide a competitive measurement of the beam neutrino energy. Right panel of figure \ref{fig2} shows the expected energy resolutions using the LDS, for beam  $\nu_e$ CC interactions in the 1GeV to 8GeV range. The energy resolution stays below 10\%, considering only the energy deposited within the active volume, and below 18\% considering the true neutrino energy. [4].

\vspace{-0.05cm}

\subsection{Proton decay searches}
DUNE will be sensitive to the $p \rightarrow K^+ \bar{\nu}$ proton decay channel among others. Due to the efficiency of LArTPCs for  particle identification, the main backgrounds for this channel are cosmic muons and neutrinos.
\vspace{-0.05cm}

\par
\begin{figure}
\centering
\includegraphics[width=.7\textwidth]{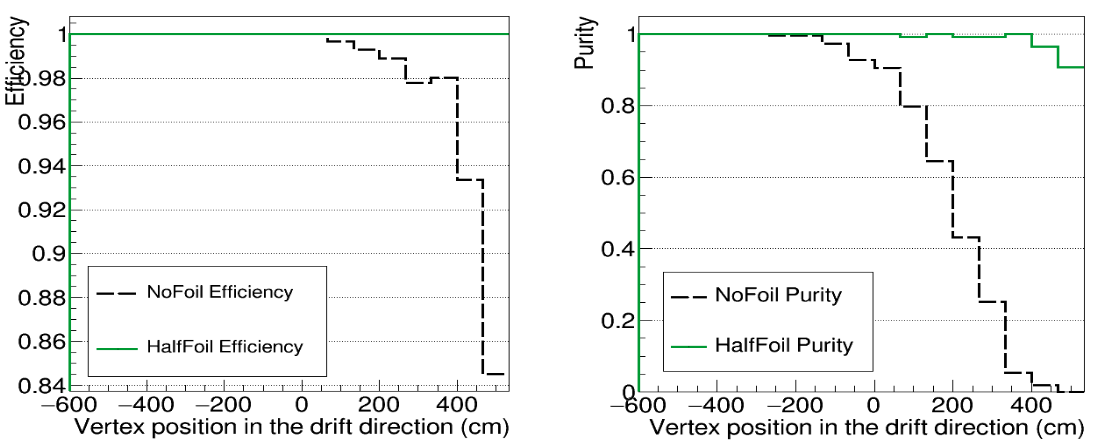}
\caption{Nucleon decay event time reconstruction efficiency (left) and purity (right) along the drift direction, in the $p \rightarrow K^+ \bar{\nu}$ channel. Two LDS designs are compared: no foil and half foils (baseline).}
\label{fig3}
\end{figure}

 In order to reject cosmic muons, a spatial veto is applied, rejecting muons close to the boundaries and effectively reducing the active volume. Within this fiducial volume, the light detection system is able to provide the event time with 100\% efficiency with a signal purity of 90\% considering the light events produced by radiological backgrounds (Figure \ref{fig3}) [4].
\par The studies presented above show that the baseline LDS design will fulfil the DUNE physics requirements. Additionally an equivalent design at a smaller scale it is being tested now at CERN in the protoDUNE DP detector.

\vspace{-0.7cm}

\acknowledgments{
The project that gave rise to these results received the support of a fellowship from ``la Caixa'' Foundation (ID 100010434). The fellowship code is LCF/BQ/IN18/11660043.}


\begin{thebibliography}{99}
\bibitem{1}
The DUNE Collaboration. (2018).
\emph{The DUNE Far Detector Interim Design Report Volume 1: Physics, Technology and Strategies},
[{\tt arXiv:1807.10334}].

\bibitem{2}
D. Belver et al. (2018).
\emph{Cryogenic R5912-20Mod Photomultiplier Tube Characterization for the ProtoDUNE Dual Phase Detector},
\emph{J. Inst.} {\bf  13 } T10006

\bibitem{3}
D. Belver et al. (2019).
\emph{A light Calibration System for the ProtoDUNE-DP detector},
\emph{J. Inst.} {\bf  14 } T04001

\bibitem{4}
The DUNE Collaboration. (2019).
\emph{TDR Physics Volume},
\emph{(in preparation)}.

\end{thebibliography}
\end{document}